\documentclass[prb,onecolumn,showpacs,preprintnumbers,amsmath,amssymb,
superscriptaddress,floatfix]{revtex4-1}
\usepackage{graphicx}
\usepackage{color}
\usepackage{dcolumn}
\usepackage{bm}
\usepackage[latin1]{inputenc}
\begin{document}
\title[Material laws and related uncommon phenomena in type-II
superconductors]{Material laws and related uncommon phenomena in the
electromagnetic response of type-II superconductors in longitudinal geometry}
\author{H. S. Ruiz}
\email{hsruizr@unizar.es}
\address{Departamento de F\'{\i}sica de la Materia Condensada \\
and Instituto de Ciencia de Materiales de Arag\'on (ICMA),\\
Universidad de Zaragoza--CSIC,
Mar\'{\i}a de Luna 3, E-50018 Zaragoza, Spain}

\author{A. Bad\'{\i}a\,--\,Maj\'os}
\address{Departamento de F\'{\i}sica de la Materia Condensada \\
and Instituto de Ciencia de Materiales de Arag\'on (ICMA),\\
Universidad de Zaragoza--CSIC,
Mar\'{\i}a de Luna 3, E-50018 Zaragoza, Spain}

\author{C. L\'opez}
\address{Departamento de Matem\'aticas, Universidad de Alcal\'a
de Henares, E-28871 Alcal\'a de Henares, Spain}
\date{\today}
\begin{abstract}

Relying on our theoretical approach for the superconducting critical state
problem in 3D magnetic field configurations, we present an exhaustive analysis
of the electrodynamic response for the so-called longitudinal transport problem
in the slab geometry. A wide set of experimental conditions have been
considered, including modulation of the applied magnetic field either
perpendicular or parallel (longitudinal) to the transport current density. The
main objective of our work  was to characterize the role of the macroscopic
material law that should properly account for the underlying mechanisms of flux
cutting and depinning. The intriguing occurrence of negative current patterns
and the enhancement of the transport current flow along the center of the
superconducting sample are reproduced as a straightforward consequence of the
magnetically induced internal anisotropy. Moreover, we show that related to a
maximal projection of the current density vector onto the local magnetic field,
a maximal transport current density occurs somewhere within the sample.  The
elusive measurement of the flux cutting threshold (critical value of such
parallel component $J_{c ||}$) is suggested on the basis of local measurements
of the transport current density. Finally, we show that a high correlation
exists between the evolution of the transport current density and the appearance
of paramagnetic peak structures in terms of the applied longitudinal magnetic
field.

\end{abstract}
\maketitle


\section{Introduction}\label{Section_1}
The high interest arisen concerning the problem of magnetic flux depinning in
type-II superconductors is markedly associated with its relevance to
technological and industrial applications achieving elevated transport currents
with no discernible energy dissipation. Thus, the distribution of vortices in a
real type-II superconductor is determined by the balance between the
electromagnetic driving forces perpendicular to the flux lines, and forces
pinning the vortices to material inhomogeneities (lattice defects, impurities,
and finite-size effects)~\cite{Rosentein_2010,Voloshin_2010,Sanchez_2010}.  Per
unit volume, this reads ${\bf J}\times{\bf B}={\bf F}_p$ (or $J_{\perp}B =
F_p$). 

On the other hand, it is well-known that various striking phenomena occur when a
transport current is applied to a nonideal type-II superconductor under the
presence of a longitudinal magnetic
field~\cite{Badia_2009,Ruiz_2010,Ruiz_2011,Blamire_2003,LeBlanc_2003,
LeBlanc_2002,Voloshin_2001,Matsushita_1998,LeBlanc_1993,LeBlanc_1991,
Voloshin_1991,Matsushita_1984,Cave_1978,Wamsley_1977,Esaki_1976,Wamsley_1972,
London_1968,LeBlanc_1966,Watanabe_1992,Blamire_1986,Boyer_1980,Gauthier_1974,
Karasik_1970,Sugahara_1970,Taylor_1967,Sekula_1963,Clem_1980_1,Clem_1980_2,
Nakayama_1972}. First, a remarkable enhancement of the critical current density
has been observed in a wide number of conventional and high temperature
superconducting
systems~\cite{Sanchez_2010,Blamire_2003,Watanabe_1992,Blamire_1986,Boyer_1980,
Gauthier_1974,Karasik_1970,Sugahara_1970,Taylor_1967,Sekula_1963,Clem_1980_1,
Clem_1980_2}. Second, an intriguing negative longitudinal electric field along
the direction of the transport current has been also
reported~\cite{LeBlanc_2003,LeBlanc_2002,Voloshin_2001,Matsushita_1998,
LeBlanc_1993,LeBlanc_1991,Voloshin_1991,Matsushita_1984,Cave_1978,Wamsley_1977,
Esaki_1976,Wamsley_1972,London_1968,LeBlanc_1966}. Such a resistive structure in
the longitudinal field geometry has been intuitively understood in terms of
helical domains, closely connected to the force-free current parallel to
flux-lines (recall that the magnetic force per unit volume is given by ${\bf
J}\times{\bf B}$)~\cite{LeBlanc_1966,Gauthier_1974,Esaki_1976,Matsushita_1998}.

Nevertheless, although several facts suggest that the vortex lattice is arranged
in a helical configuration, perhaps close to the force-free arrangement (${\bf
J}\|{\bf B}$), the arising voltage cannot be straightforwardly described as a
critical flux-flow voltage if one is to be consistent with the experimental
reports~\cite{Nakayama_1972,LeBlanc_1991}. On the other hand, although several
proposals have been done in terms of the crossing and recombination of adjacent
nonparalell flux lines, the so-called flux line cutting phenomenon has been
basically recognized as the physical mechanism by which the longitudinal voltage
is
produced~\cite{Esaki_1976,Wamsley_1972,London_1968,Clem_1980_1,Clem_1980_2,
Brandt_1980,Campbell_1972,Clem_1975}. Outstandingly, remarkable numerical and
conceptual difficulties related to the implementation of the above picture of
the local dynamics for the transport current in realistic longitudinal
geometries have lead to unfortunate omissions of the phenomenon in practical
calculations related to the applications of type-II materials. 

Recently, we have shown that our general critical state theory
(Ref.~\cite{Badia_2009,Ruiz_2010}), is able to describe on a quantitative basis
the anomalous features involved in the local electrodynamics of a current
carrying superconducting sample subjected to variations of both longitudinal and
transverse magnetic fields (Ref.~\cite{Ruiz_2011}). In brief, our work is
developed at a phenomenological level which allows to deal with 3D-magnetic
field configurations and nontrivial constraint conditions defining anisotropy
effects on the superconducting material law. We have introduced a geometrical
description for the constraints on the critical current density in terms of a
closed region $\Delta_{\textbf{r}}$, such that the physically admissible states
are given by the condition ${\bf J}\in\Delta_{\textbf{r}}$. When the physical
limitations concerning the microscopic phenomena of flux depinning and cutting
are considered to be independent, the region $\Delta_{\textbf{r}}$ becomes a
cylinder, with a longitudinal rectangular section of size $2J_{c\perp}\times
2J_{c\|}$. This {\em ansatz}\cite{Badia_2009,Ruiz_2010} can be identified with
the so-called {\em double critical state model} in type-II superconductivity.
Hereafter, a related anisotropy parameter, that we will name after {\em
bandwidth} $\chi\equiv J_{c\|}/J_{c\perp}$ will be used.

Preliminar work on the intriguing effects associated to the longitudinal
transport problem was presented in Ref.~\cite{Ruiz_2011}. Here, supported by
numerical simulations that cover an extensive set of experimental conditions, we
put forward a much more complete physical scenario. Thus, we will show that the
striking existence of negative flow domains, local and global paramagnetic
structures, emergence of peak-like structures in both the critical current
density and the longitudinal magnetic moment, as well as the compression of the
transport current in type-II superconductors under parallel magnetic fields, are
all predicted by our general critical state model. In addition, we shall
introduce some ideas that could be applied for the determination of the flux
cutting threshold from local measurements of the current density flowing along
specific layers of the superconducting sample, as correlated to the behavior of
the magnetic moment components.

The paper is organized as follows. In Sec.~\ref{Section_2}, the physical
background for the critical state concept in longitudinal geometries under
applied 3D magnetic fields is introduced (Sec.~\ref{SubSection_2_1}). Then, the
underlying approximations and the variational statement for the longitudinal
transport problem in superconducting slabs are both described in detail
(Sec.~\ref{SubSection_2_2}). This encloses the time-discretized description of
the magnetic field penetration process, and the local dynamics of the transport
current in an arbitrary magnetization process. The theory is then applied for a
set of experimental configurations and a collection of material laws as regards
the interplay of the flux deppining and cutting mechanisms
(Section~\ref{Section_3}). On the one hand, the infinite bandwidth model (or
T-state model) with $J_{c ||}\gg J_{c \perp}$ is assumed
(Sec.~\ref{SubSection_3_1}). As the necessity of including a model with a well
defined value for the threshold $J_{c||}$ will become apparent, in
sections~\ref{SubSection_3_2} and \ref{SubSection_3_3} a more general
description is presented as function of the anisotropy effects on the material
law. These effects will be investigated in terms of the range of parameter
values: $\chi=1,2,3,4$. The procedure will reveal both local and global
properties of the magnetic moment and the transport current flow indicating a
possible reconstruction scheme of the underlying material law.
Sec.~\ref{Section_4} is devoted to summarize the main findings of our work.
%
%
%
\section{Longitudinal transport problem in the double critical state
regime}~\label{Section_2}
This section is devoted to introduce the theoretical background that justifies
our variational statement as an appropriate tool which allows to implement a
wide class of material laws in the investigation of the electromagnetic response
of superconducting samples subjected to 3D magnetic field configurations and
transport currents. The characteristic mathematical equations for the critical
state problem are introduced in terms of a variational statement that is applied
within the infinite slab geometry. The relevance of the parallel and
perpendicular critical current limitations will be demonstrated.
%
%
%
\subsection{General statements}~\label{SubSection_2_1}
The dynamics of the local magnetic field and transport current profiles in
type-II superconductors is commonly obtained within the critical state model
framework. In the basic formulation (Bean's
model,~\cite{Bean_1962,Bean_1964,Bean_1970}) one considers that the magnetic
flux lines, when driven by a macroscopic current density, will either penetrate
or exit across the superconducting surface. In more detail, when the driving
forces (local value of ${\bf J}\times{\bf B}$) overcome the pinning force in a
local section of the superconducting sample, the system of vortices rearranges
itself into a new metastable state such that the vortex lattice is pinned again
when the equilibrium at the boundary is reestablished. Notice that this physical
mechanism relates to the component of ${\bf J}$ perpendicular to ${\bf B}$.
Since the displacement of the flux-lines takes place with high associated
resistivity, the system quickly adjusts itself to successive equilibrium states
so as to avoid the high resistive losses. On the other hand, reversible energy
terms related to the equilibrium properties of the vortex lattice are usually
neglected (justified by the range of interest for the local magnetic fields
$H_{c1}\ll H\ll H_{c2}$), and the linear relation ${\bf B}=\mu_{0}{\bf H}$ is
assumed. 

For the particular case that the transport current is applied along the
direction of the vortex lattice, rotation of vortices is induced by the
component of the current density parallel to the magnetic field. Subsequent
cross-joining and cutting will occur if the angle between neighboring vortices
overcomes some critical value. Thus, in general, for nontrivial configurations,
and related to the threshold values for the current density that trigger the
displacement of vortex lines in some sense, a critical state may be defined as a
given configuration which is able to withstand a critical current defined by
${\bf J}_{c}={\bf J}_{c ||}+{\bf J}_{c \perp}$, as long as neither threshold
(${\bf J }_{c ||}$ or ${\bf J }_{c\perp}$) is
exceeded~\cite{Badia_2009,Ruiz_2010,Ruiz_2011,Clem_1980_1,Clem_1980_2,Clem_1975,
Clem_1977}.

One must mention that, data from a wide number of experiments have shown that in
many cases $J_{c ||}$ and $J_{c\perp}$ can be of the same order of
magnitude~\cite{Boyer_1977,Boyer_1980,Fillion_1979,Cave_1982,LeBlanc_1984,
LeBlanc_1991,LeBlanc_1993,LeBlanc_2003,LeBlanc_2002,Voloshin_2001}. Then, in
order to explain the accompanying phenomena, both thresholds have to be included
in the theory.  In these cases, the local current voltage characteristic,
connecting the local electric field ${\bf E}$ and current density $\textbf{J}$,
is strongly anisotropic with respect to the direction of the magnetic induction
${\bf B}$. This fact introduces strong difficulties when trying to find an
analytical solution for the \textit{critical-state equation} $\nabla\times{\bf
H}=\textbf{J}_{c}$ (or 0). Our formulation can somehow {\em bypass} such
difficulties by treating the electric field as a derived quantity that may be
obtained {\em a posteriori}.

Thus, the critical-state equation can be posed in numerical terms, assuming an
evolutionary discretization scheme. Let us suppose that ${\bf H}_{l}$ stands for
the local magnetic field intensity at the time layer $l\delta t$, and that the
current density profiles relate to some \textit{magnetic diffusion} process that
takes place when the local conditions for the material law $J_{\parallel}\leq
J_{c\parallel},J_{\perp}\leq J_{c\perp}$ are violated. The successive field
penetration profiles within the superconductor may be obtained by the
finite-difference expression of Faraday's law,
%
%
\begin{eqnarray}
\mu_{0}\frac{\textbf{H}_{l+1}-\textbf{H}_{l}}{\delta
t}=-\nabla\times\textbf{E}=-\nabla\times\left[
\rho(\nabla\times\textbf{H}_{l+1})\right]\, .
\label{Eq_1}
\end{eqnarray} 
Here, $\rho ({\bf J})$ plays the role of a nonlinear and nonscalar resistivity
that should properly incorporate the physics of the threshold and dissipation
mechanisms mentioned above.
Notice that the local profile $\textbf{H}_{l+1}$ can be solved in terms of the
previous field distribution $H_{l}$ and the boundary conditions at time layer
$(l+1)\delta t$. The initial condition fulfills Ampere's law
$\nabla\times\textbf{H}_{l}=\textbf{J}_{l}$ as well as
$\nabla\cdot\textbf{H}_{l}=0$ and $\nabla\cdot\textbf{J}_{l}=0$. One possibility
for making the integration of this system affordable is to find an equivalent
variational statement. Thus, one can avoid the integration of these set of
differential equations by \textit{inversion} of the Euler-Lagrange equations
%
%
\begin{eqnarray}
\textbf{J}_{l+1}-\nabla\times\textbf{H}_{l+1}=0\, ,
\label{Eq_2}
\end{eqnarray} 
and
%
%
\begin{eqnarray}
\nabla\times\textbf{p}_{l}+\textbf{H}_{l+1}-\textbf{H}_{l}=0\, ,
\label{Eq_3}
\end{eqnarray} 
for arbitrary variations of the time-discretized local magnetic field,
$\textbf{H}_{l}$, that plays the role of the unknown. On the other hand, we have
introduced the Lagrange multiplier,  $\textbf{p}_{l}$, which can basically
identified with the electric field of the problem. The related Lagrange density
(over whole space) is
%
%
\begin{eqnarray}
L=\frac{1}{2}|\textbf{H}_{l+1}-\textbf{H}_{l}|^{2}+\textbf{p}
\cdot(\nabla\times\textbf{H}_{l+1}-\textbf{J}_{l+1})\, .
\label{Eq_4}
\end{eqnarray} 
As indicated above, minimization under the constraint
$\nabla\times\textbf{H}_{l+1}=\textbf{J}_{l+1}$ (Amp\`ere's law) allows to
reduce the number of variables in the statement of the problem, skipping the
explicit inclusion of $\bf E$.

For practical purposes, we stress that the above equations may be transformed so
as to get a vector potential formulation by means the condition
%
%
\begin{eqnarray}
\textbf{H}_{l+1}-\textbf{H}_{l}=\mu_{0}^{-1}\nabla\times(\textbf{A}_{l+1}
-\textbf{A}_{l}) \, .
\label{Eq_5}
\end{eqnarray} 
Thus, by using standard electromagnetic manipulations, one may show that the
functional to be minimized can take the following  form in terms of the current
density
%
%
\begin{eqnarray}
{\cal F}~[\textbf{J}_{l+1}\in\Delta_{\textbf{r}}] \equiv&& \frac
{8\pi}{\mu_0}\int_{\Omega} \Delta \textbf{A}_{0}\cdot {\bf J}_{l+1}d{\bf r}
\nonumber\\
&&+\int\int_{\Omega\times\Omega}
\frac{\textbf{J}_{l+1}'\cdot[\textbf{J}_{l+1}-2\textbf{J}_{l}]}{|\textbf{r}
-\textbf{r}'|}d\textbf{r}d\textbf{r}' \nonumber\\ \; 
\label{Eq_6}
\end{eqnarray}
It must be emphasized that Eq.(\ref{Eq_6}) can be applied for any shape of the
superconducting volume $\Omega$ as well as for any general restriction (material
law) for the current density $\textbf{J}_{l+1}\in\Delta_{\textbf{r}}$.  It is
also to be noticed that solenoidality ($\nabla\cdot\textbf{J}=0$) has to be
imposed so as to be consistent with charge conservation in quasi-steady regime.
%
%
%
%

\subsection{Variational statement in slab geometry}~\label{SubSection_2_2}
\begin{figure}
\begin{center}
{\includegraphics[width=0.6\textwidth]{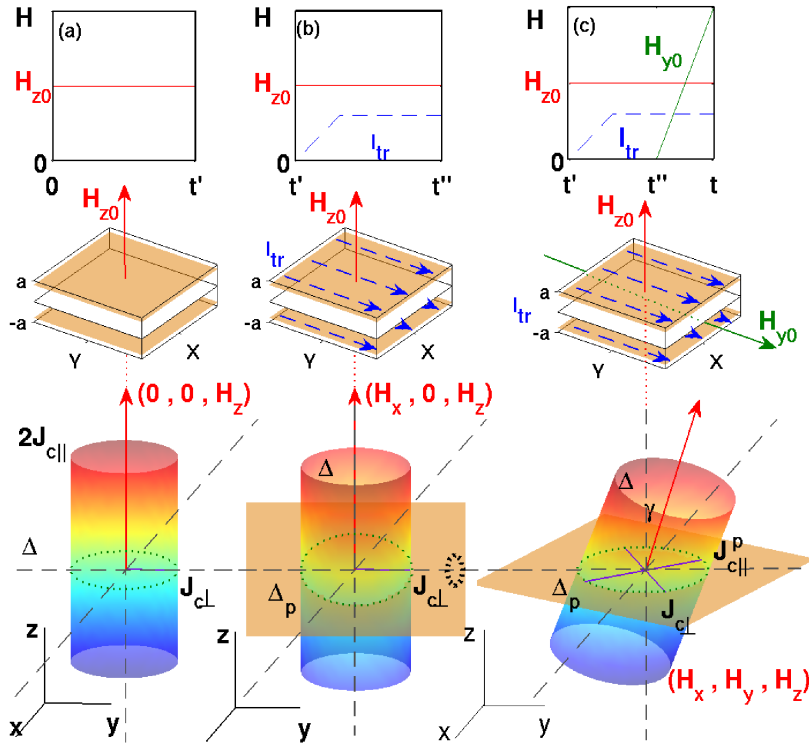}}
\caption{(Color online) Scheme of the time dependence of the magnetic process
considered in this work. (a) Left panel: an external magnetic field $H_{z0}$ is
applied perpendicular to the external surface of a superconducting slab of
thickness $2a$. (b) Middle: the slab is later subjected to a transport current
along the $y$-axis. This generates a local magnetic field component $H_{x}$. (c)
Right panel: at $t=t''$ an external magnetic field $H_{y}$ parallel to the
transport of current direction is switched on. The lower row of each panel
depicts the region $\Delta$ that constrains the flow of electrical current, as
related to the critical state material law. $\Delta$ is a cylinder with axis
parallel to the local magnetic field, with radius $J_{c\perp}$ and length
$2J_{c\|}$. The components of ${\bf J}$ within the sample's plane are indicated
for each case}\label{Fig_1}
\end{center}
\end{figure}

In this subsection, we derive a specific variational formulation for the
longitudinal transport problem in superconducting slabs for eventual 3D-field
configurations (Fig.~\ref{Fig_1}), i.e., both in plane and transverse local
magnetic field components emerge as derived effects of the external sources
(transport currents and/or external magnetic sources).

We shall consider the time evolution of magnetic profiles $\textbf{H}_{l+1}(z)$
within an infinite superconducting slab (of thickness $2a$), cooled under the
assumption of an initial state defined by a uniform vortex lattice perpendicular
to the external surfaces, i.e., a constant magnetic field $H_{z0}$. Then, the
constraint for the current density  (material law ${\bf
J}\in\Delta_{\textbf{r}}$) corresponding to the limitations on the parallel and
perpendicular directions mentioned above may be visualized by a finite cylinder
with its axis parallel to the local magnetic field $H_{z}$ (Fig.~\ref{Fig_1}a).
When a transport current is injected along the superconducting slab in the
direction of y-axis, a magnetic field component $H_{x}$ appears, inducing a
rotation of the critical current region $\Delta_{\textbf{r}}$ (see
Fig.~\ref{Fig_1}b). Finally, if a third magnetic field component $H_{y}$ is
switched, a new rotation of the region occurs (Fig.~\ref{Fig_1}c). Notice that,
by symmetry, the current density is confined to the $XY$ plane, i.e.: ${\bf
J}=(J_{x}(z),J_{y}(z),0)$. In particular, this means that in practice one should
impose the restriction that ${\bf J}$ belongs to the projection of the critical
current region onto the plane (${\bf J}\in\Delta_p$) and that
$\nabla\cdot\textbf{J}\equiv 0$.

At this point, we must mention that for numerical convenience, the material law
is not strictly used in the form of a cylinder, but smoothly reshaped as a {\em
superellipsoid}~\cite{Badia_2009,Ruiz_2010} by means of the relation

%
%
\begin{equation}
\left(\frac{J_{||}}{J_{c ||}}\right)^{2n}+\left(\frac{J_{\perp}}{J_{c
\perp}}\right)^{2n}\leq 1\, .
\label{Eq_7}
\end{equation}
The critical current region $\Delta_{\textbf{r}}$ will be characterized in terms
of its bandwidth $\chi=|J_{c ||}/J_{c\perp}|$ and the superelliptic index $n$.

We call the readers' attention that two analytical approaches for the slab
geometry may be found in the literature for extreme situations. The first one
was introduced by Brandt and Mikitik in Ref.~\cite{Brandt_2007} for the regime
of strong pinning with very weak longitudinal current conditions, i.e., $H_{z}$
must be very high as compared to $H_{xy}(a)$ (then $J_{\|}\ll J_{c\|}$). On the
other hand, the opposite limit ($H_{z}\to 0$) was recently considered in
Ref.~\cite{Ruiz_2011}. Here, and based on the numerical resolution of the
variational statement, a complete tour along the whole set of values for the
perpendicular field will be presented.

\begin{figure}[t]
\begin{center}
{\includegraphics[width=0.6\textwidth]{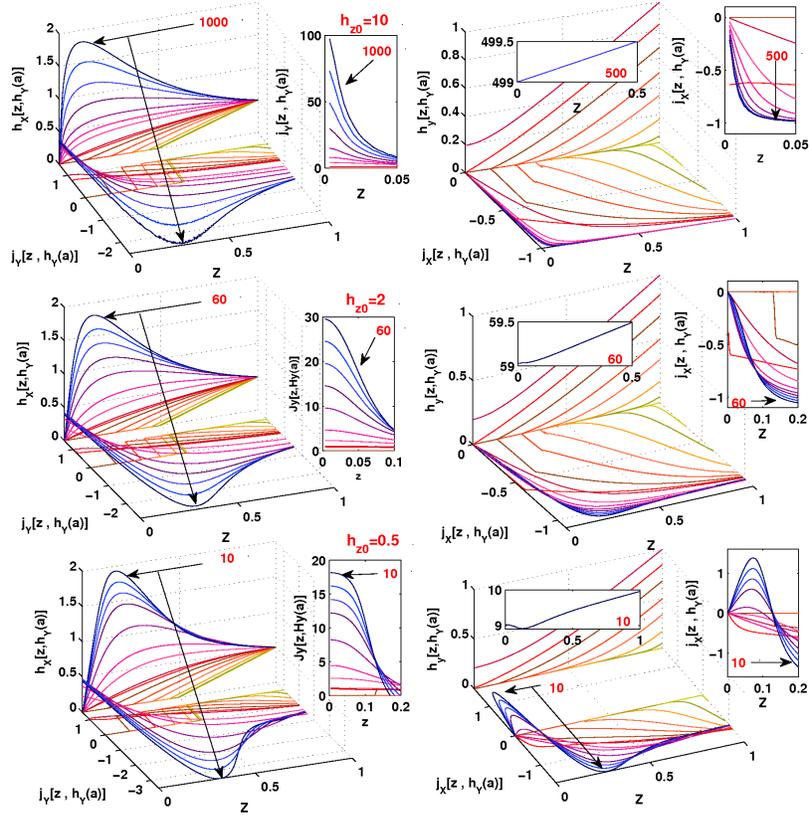}}
\caption{(Color online) Profiles of the magnetic field components
$h_{x}[z,h_{y}(a)]$ and $h_{y}[z,h_{y}(a)]$, and the corresponding
current-density profiles $j_{y}[z,h_{y}(a)]$ and  $j_{x}[z,h_{y}(a)]$ for the
\textit{T}-state model and perpendicular magnetic field components $h_{z0}=10$
(top), $h_{z0}=2$ (middle) and $h_{z0}=0.5$ (bottom). The different curves
correspond to the following sets of values for the longitudinal field at the
surface: (i) top row: $h_{y}(a)=0.005, 0.050, 0.170, 0.340, 0.500, 0.680, 0.845,
1.0, 40.0,80.0, 150.0, 300.0, 500.0, 750.0$, $1000$, (ii) middle row:
$h_{y}(a)=0.0050.050,0.170,0.340,0.500,0.680,0.845,1.0,5.0,10.0$,
$20.0,30.0,40.0,50.0,60.0$, (iii) bottom row: $h_{y}(a)=0.001,0.050,
0.170,0.340,0.500$, $0.680, 0.845, 1.0, 2.0, 3.0,5.0,7.0, 8.0, 9.0, 10.0$.}
\label{Fig_2}
\end{center}
\end{figure}

In order to simplify the mathematical statements we shall normalize the
electrodynamic quantities by defining $\textbf{h}\equiv\textbf{H}/J_{c\perp}a$,
$\textbf{j}\equiv\textbf{J}/J_{c\perp}$, and $\texttt{z}\equiv z/a$. In turn,
our problem will be described in terms of $N_{s}$ discretized layers, each one
characterized by a current density function
$\textbf{j}(z_{i})=\textbf{j}_{x}(z_{i})+\textbf{j}_{y}(z_{i})$ distributed
along $|z_{i}|\leq N_{i} a/N_{s}$. For further consideration, notice that the
value of the transport current assumed in this work, i.e.: penetration to half
thickness, imposes the boundary condition $h_{x}(a) = 0.5$ (Fig.~\ref{Fig_2}).

In our discretized description, and in terms of Ampere's law $h_{x}(z_{i})$ will
be evaluated from
%
%
\begin{equation}
h_{x}(z_{i})=\delta\sum_{j<i} \frac{2 j_{y}(z_{j})-j_{y}(z_{i})}{2}\, .
\label{Eq_8}
\end{equation}

Similarly, the local profiles for the longitudinal magnetic field component
$h_{y}(z_{i})$ can be obtained from
%
%
\begin{equation}
h_{y}(z_{i})=\delta\sum_{j>i} \frac{2 j_{x}(z_{j})-j_{x}(z_{i})}{2}\, ,
\label{Eq_9}
\end{equation}
with $\delta\equiv a/N_s$ the thickness of each layer over the local plane
$z_{i}$. 

Then, for the magnetic process sketched in Fig.~\ref{Fig_1}c, Eq.(\ref{Eq_6})
takes the following form in terms of discrete variables
%
%
\begin{eqnarray}
{\cal F}(t=l+1)&&=
\frac{1}{2}\sum_{i,j}I_{i,l+1}^{x}M_{ij}^{x}I_{j,l+1}^{x}-\sum_{i,j}I_{i,l}^{x}
M_{i,j}^{x}I_{j,l+1}^{x}\nonumber \\ 
&&
+\frac{1}{2}\sum_{i,j}I_{i,l+1}^{y}M_{ij}^{x}I_{j,l+1}^{y}-\sum_{i,j}I_{i,l}^{y}
M_{i,j}^{y}I_{j,l+1}^{y}\nonumber \\
&& +\sum_{i}I_{i,l+1}^{y}(i-1/2)(h_{l+1}^{y}(a)-h_{l}^{y}(a))\, .
\label{Eq_10}
\end{eqnarray}
In this expression we have introduced the sheet currents $I_{i,l+1}^{x}\equiv
\delta j_{x}(z_{i}, t= l+1)$ and $I_{i,l+1}^{y}\equiv \delta j_{y}(z_{i}, t=
l+1)$, and their {\em mutual inductance} coefficients $M_{ij}^{x,y}$. 
${\cal F}$ has to be minimized along the different time steps ($l=1,2,\dots$) in
the magnetic process under consideration.
On the other hand, one can show that the parallel and perpendicular {\em
projections} of the sheet current components are given by
%
%
\begin{eqnarray}
I_{\perp}^{2} &=&
(1-h_{x,i}^{2})I_{x,i}^{2}+(1-h_{y,i}^{2})I_{y,i}^{2}-2h_{x,i}h_{y,i}\,I_{x,i}I_
{y,i} \, ,
\nonumber\\
I_{\parallel}^{2} &=&
h_{x,i}^{2}\,I_{x,i}^{2}+h_{y,i}^{2}\,I_{y,i}^{2}+2h_{x,i}h_{y,i}\,I_{x,i}I_{y,i
} \, .
\label{Eq_11}
\end{eqnarray}
These components have to be constrained according to the material law in
Eq.(\ref{Eq_7}).
Further, for the transport problem, one has to consider the external constraint:
%
%
\begin{eqnarray}
\sum_{i}I_{y,i}=I_{tr} \, .
\label{Eq_12}
\end{eqnarray}

The theoretical framework becomes closed by the following expressions for the
mutual inductance coupling elements\cite{Badia_2009,Ruiz_2010}
%
%
\begin{eqnarray}
M_{i,j}^{x}\equiv 1+2\left[min\{i,j\} \right]\nonumber \, , \\ \nonumber \\
M_{i,i}^{x}\equiv 2\left(\frac{1}{4}+i-1 \right)\nonumber \, , \\ \nonumber \\
M_{i,j}^{y}\equiv 1+2\left[N_{s}-max\{i,j\} \right]\nonumber \, , \\ \nonumber
\\
M_{i,i}^{y}\equiv 2\left(\frac{1}{4}+N_{s}-i\right) \, .
\label{Eq_13}
\end{eqnarray}

In summary, the magnetic response of the superconductor is characterized by a
collection of discretized current elements for the planar sheets 
$(j_{xi},j_{yi})$ at the time steps $l+1=1,2,3,...$. Then, the magnetic field
profiles may be obtained by means of Eqs.(\ref{Eq_8}) and (\ref{Eq_9}), and  the
sample's magnetic moment per unit area from
%
%
\begin{eqnarray}
{\bf M}_{l+1}=\sum_{i} {\bf z}\times{\bf j}_{i} \, ,
\label{Eq_14}
\end{eqnarray}
where a factor of 2 has been introduced related to the contribution of the
U-turns at infinity for each circuit.
%

%

%
%
%


\section{Influence of the magnetic anisotropy in the critical
current}~\label{Section_3}
In the previous sections we have visualized the double critical state model as a
phenomenological approach which can be formulated by means of a variational
problem with physical constraints. Here, based on the above mentioned
theoretical statements for the longitudinal transport current problem, we will
show the theoretical predictions for the magnetization process outlined in
Fig.~\ref{Fig_1}c as $h_{y}(a)\equiv H_{y0}$ is increased. In addition, several
initial states $h_{z0}$ will be focused on. Moreover, we shall concentrate on
the effect of the flux cutting boundary ($j_{c ||}$) considering several
conditions for the material law. Two extreme cases ($\chi = 1$ and
$\chi\to\infty$) will be considered first (Sections \ref{SubSection_3_1} and
\ref{SubSection_3_2}), and the range in between at a second step ($\chi=2$, 3,
and 4 in Section~\ref{SubSection_3_3}). Remarkably, our procedure will reveal
the fingerprints of the cutting and depinning mechanism, thus being a
theoretical pathway for the reconstruction of the material law, represented by
the proper region $\Delta_{\textbf{r}}$.

Henceforth, we shall use the simplified notation T or CT$\chi$ as regards to the
infinite bandwidth model (T by transport) or double critical state model (CT by
cutting and transport) with anisotropy $\chi=|j_{c ||}/j_{c \perp}|$.

%
%
%
%
\subsection{T-states for the longitudinal transport
configurations}~\label{SubSection_3_1}

\subsubsection{Field and current density penetration
profiles}~\label{SubsubSection_3_1_1}
Fig.~\ref{Fig_2} shows the magnetic field profiles and the induced currents for
three different initial conditions, i.e., $h_{z0}=10$, $h_{z0}=2$, and
$h_{z0}=0.5$, all of them under assumption of the T-state model. The initial
state for the transport current condition ($I_{tr}=J_{c \perp}a/2$) establishes
the initial transport profile $j_{y}\{0\leq z<a/2\}=0$ and $j_{y}\{a/2\leq z\leq
a\}=1$. As the transport current is no longer modified, the condition
$h_{x}(a)=0.5$ can be applied in what follows. On the other hand, by symmetry,
one has to impose the condition $h_{x}(0)=0$ at the center of the slab.

\begin{figure}
\begin{center}
{\includegraphics[width=0.6\textwidth]{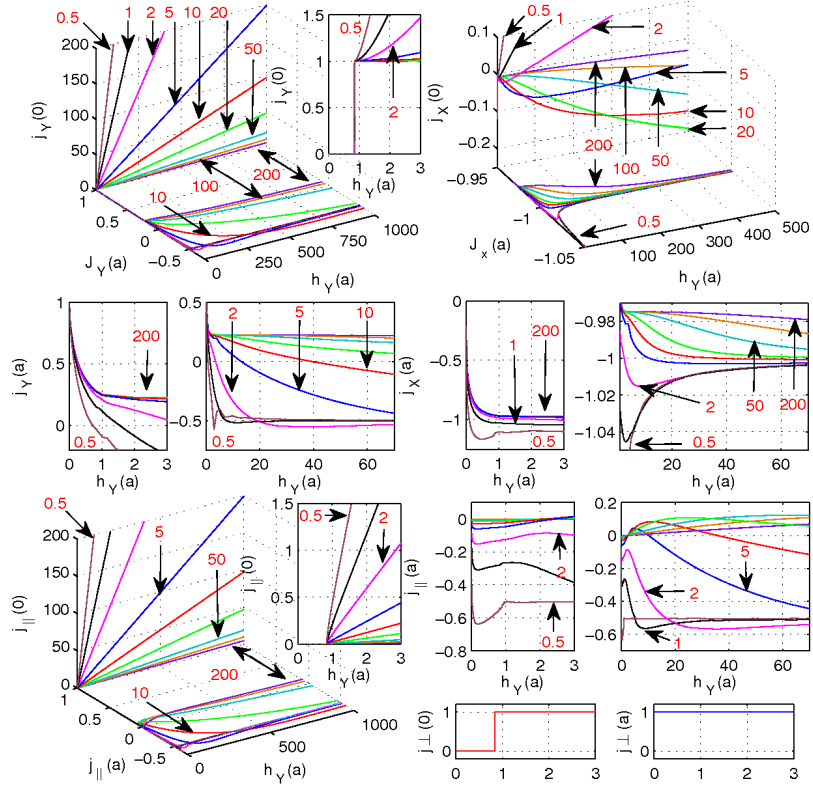}}
\caption{(Color online) Evolution of the local current density as a function of
the applied longitudinal magnetic field $h_{y0}=h_{y}(a)$ along the central and
external sheets of the slab. The results are shown for the T-state model
($J_{c||}\to\infty ~\&~ J_{c\perp}=1.0$). Top: the components $j_{y}$ and
$j_{x}$ at $(z=0)$ and $(z=a)$. Middle: details of the above behavior. Bottom:
behavior of the parallel and perpendicular components of $\bf j$ in the same
conditions as above. The different curves correspond to the values of the
perpendicular magnetic field given by $h_{z0}= 1, 2, 5, 10, 20, 50, 100, 200$
(all plots having the same color scale).}
\label{Fig_3}
\end{center}
\end{figure}

Then, when the external magnetic field $h_{y}(a)$ is linearly increased from
$h_{y}(a)=0$, a current density $j_{x}$ is induced from the superconducting
surface as an effect of the Faraday's law.  Simultaneously, the local component
of the magnetic field $h_{x}(z)$ increases monotonically following two
continuous stages fulfilling the aforementioned boundary conditions. First, the
superconducting sample is fully penetrated by the transport current when
$h_{y}^{\star}(a)=0.845\pm0.003$ and eventually, the condition $j_{y}(0)=1$ is
reached as soon as $h_{y}(a)\rightarrow 0.860$. 
We notice that the value of $h_{y}^{\star}(a)$ for the full penetration profile
is basically independent of $h_{z0}$, at least to the numerical precision of our
numerical calculations. This agrees with the analytical solution of
Ref.~\cite{Ruiz_2011}. Second, a remarkable enhancement of the transport current
density occurs around the center of slab as $h_{y}(a)$ increases over
$h_{y}^{\star}(a)$. Furthermore, an eventual negative current density appears
shielding the positive transport current around the center of slab. In more
detail, notice that the appearance of negative current flow is enhanced when the
magnetic component $h_{z0}$ is decreased (Fig.~\ref{Fig_3}). 

Outstandingly, profiles of magnetic field reentry (paramagnetism in the
component $h_{y}$ around center of slab) are obtained for $h_{z0}\lesssim 1$
under relatively low applied magnetic fields $h_{y}(a)$ (see Fig.\ref{Fig_2}). 

Another remarkable property is that, for the range of values
$h_{z}(0)<h_{y}^{\star}(a)$ negative surface current appears even for the
partial penetration regime, e.g., for $h_{z0}=0.5$ one has $j_{y}(a)<0$ for
$h_{y}(a)>0.722$. Recall that, in Ref.~\cite{Ruiz_2011} we have analytically
shown that both effects, local paramagnetism and negative current zones are also
predicted in the limiting case $h_{z0}=0$ . Along this line, as a general rule,
we can conclude that the smaller the value of $h_{z0}$, the sooner the surface
of negative transport current and even paramagnetic local effects appear. 
%
%

%


\subsubsection{Analysis of the current density
behavior}~\label{SubsubSection_3_1_2}
Fig.~\ref{Fig_3} displays the evolution of the current density vector as a
function of the longitudinal magnetic field $h_{y}(a)$. Note that we focus on
the specific values at the superconducting surface ($z=a$) and at the center of
the superconducting slab ($z=0$). 

In order to understand the physical mechanisms responsible for the observed
behavior, it will be useful to consider the following representation of the
current density components. We identify the following decomposition of the
vector {\bf J}
%
%
\begin{equation}
{\bf J}={\bf J}_{\parallel}+{\bf J}_{\perp\alpha}+{\bf J}_{\perp{\theta}}\, ,
\label{Eq_15}
\end{equation}
that assumes a polar axes representation, with the parallel, azimuth and polar
components of ${\bf J}$ defined in terms of the magnetic field direction. The
following expressions are obtained for such components, when a cartesian
coordinate system is used.

(i) The current density parallel to $\textbf{\^{H}}$ or so-called cutting
current component $J_{||}$:
%
%
\begin{eqnarray}
J_{||}=\frac{H_{x}J_{x}+H_{y}J_{y}}{H}
 \, .
\label{Eq_16}
\end{eqnarray}

(ii) The component of ${\bf J}$ perpendicular to the plane defined by the
vectors $\textbf{\^{z}}$ and $\textbf{\^{H}}$ or so-called azimuthal current
component
%
%
\begin{eqnarray}
J_{\perp \alpha}=\frac{-H_{y}J_{x}+H_{x}J_{y}}{(H_{x}^{2}+H_{y}^{2})^{1/2}}
 \, ,
\label{Eq_17}
\end{eqnarray}

(iii) the component of ${\bf J}$ perpendicular to $\textbf{\^{H}}$ and contained
in the plane defined by the vectors $\textbf{\^{z}}$ and $\textbf{\^{H}}$ or
so-called polar current component $J_{\perp \theta}$:
%
%
\begin{eqnarray}
J_{\perp
\theta}=\frac{H_{z}(H_{x}J_{x}+H_{y}J_{y})}{H(H_{x}^{2}+H_{y}^{2})^{1/2}}\, .
\label{Eq_18}
\end{eqnarray}

Within the T-state model, the behavior of ${\bf J}$ is straightforwardly
explained: the unbounded parallel current density allows unconstrained rotations
for the flux lines as the applied magnetic field increases. In particular, this
leads to negative values of $j_{y}(a)$ (slope of $h_{x}(a)$), simultaneous to
high $j_{y}(0)$ (slope of $h_{x}(0)$). We call the readers' attention that
negative values of the transport current are favored by smaller and smaller
values of the field component perpendicular to the surface of the sample
$h_{z0}$.

A property to be noticed in Fig.~\ref{Fig_3} is that , at the center of slab the
flux line dynamics is mainly governed by the longitudinal transport current
density $j_{y}(0)$. The basic idea is that for moderate values of $h_z$, when
$h_{y}$ increases $j_{y}$ practically becomes $j_{\parallel}$. As this component
is unconstrained, it grows indefinitely at the center.

\subsubsection{Magnetic moment of the sample}
\begin{figure}
\begin{center}
{\includegraphics[width=0.6\textwidth]{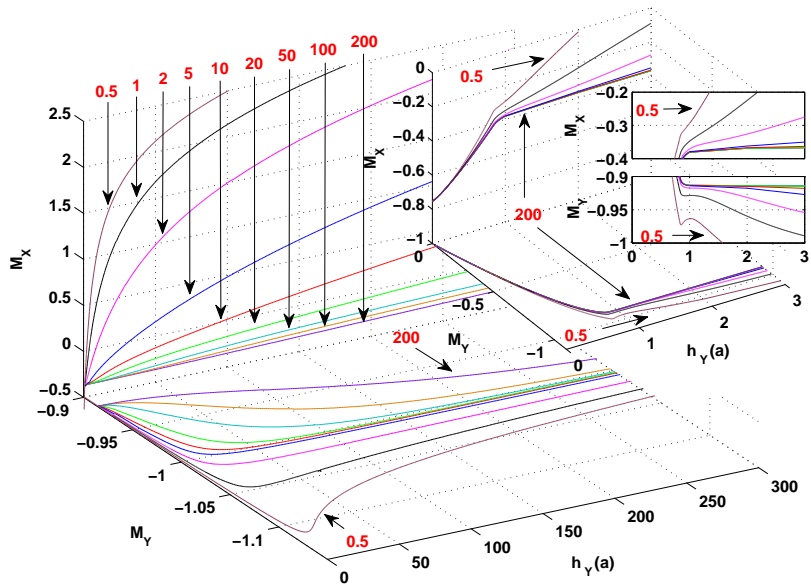}}
\caption{(Color online) The magnetic moment $(M_{x},M_{y})$ of the slab as a
function of the applied magnetic field component $h_{y}(a)$ for the T-state
model. The curves are labelled according to the perpendicular magnetic field
component $h_{z0}= 1, 2, 5, 10, 20, 50, 100, 200$, the same color scale applies
to all the plots.}
\label{Fig_4}
\end{center}
\end{figure}

Let us now concentrate on the magnetostatic properties by means of the
\textit{global} sample's magnetization curve \textbf{M}(\textbf{H}). Thus, we
have calculated \textbf{M} as a function of the longitudinal magnetic field
$h_{y}(a)$. Fig.~\ref{Fig_4} displays the magnetic moment components
$M_{x}(h_{y0})$ and $M_{y}(h_{y0})$ in units of $J_{c\perp}a^{2}$. Notice first
that within the partial penetration regime ($h_{y}(a)\leqslant
h_{y}^{\star}(a)$) the magnetic moment components are almost independent of the
transversal magnetic field $h_{z0}$ (at least for non small values of this
quantity). On the contrary, when $h_{z0}<1$ and the patterns of negative current
even occur before of the full penetrated state, magnetization slightly
increases. This is accompanied by faint field {\em reentry} effects that are
also shown in the figure. Furthermore, as the threshold cutting current $j_{c
||}$ is unbounded for the T-state model, the magnetic moment $M_{x}$ always
increases as related to the diverging behavior of $j_{y}(0)$.

%

%
%

We must emphasize that the unbounded behavior for the parallel current density
assumed above (that leads to the prediction of arbitrarily high values of the
transport current density) must be physically reconsidered. Thus, the trend of
the magnetic moment $M_{x}$ and also the unbounded longitudinal current density
$j_{y}$ disagree with the experimental evidences recollected in
Refs.~\cite{Sekula_1963,LeBlanc_1966,Karasik_1970,Sugahara_1970,Taylor_1967,
Gauthier_1974,LeBlanc_2003,LeBlanc_2002,Voloshin_2001,Esaki_1976}.  In addition,
in Fig.~\ref{Fig_3} one can notice that, as soon as the flow of negative current
along the superconducting surface is reached, it never disappears
notwithstanding the longitudinal magnetic field remains increasing. By contrast,
the disappearance of the patterns of superficial negative current were detected
in
Refs.~\cite{Voloshin_1991,LeBlanc_1966,Esaki_1976,Matsushita_1998,
Matsushita_1984}. These observations have lead to consider $J_{c\parallel}$~{\em
--bounded} descriptions as satisfactory solutions of the peculiar phenomena
involved on the longitudinal transport current
problem~\cite{LeBlanc_2003,LeBlanc_2002,Matsushita_1998,Matsushita_1984}. Rather
recent experimental data on high temperature superconductors~\cite{Clem_2011}
also indicate that physical bounds are to be considered for both components of
the critical current.

On the other hand, the method described in this work suits the necessity of
dealing with a physically acceptable description of both local and global issues
about the electromagnetic quantities involved on the longitudinal transport
current problem~\cite{Ruiz_2011}. More realistic models for the material law are
presented below.

%
%
\subsection{CT1 states as a general approach}~\label{SubSection_3_2}

%
\begin{figure}[b]
\begin{center}
{\includegraphics[width=0.6\textwidth]{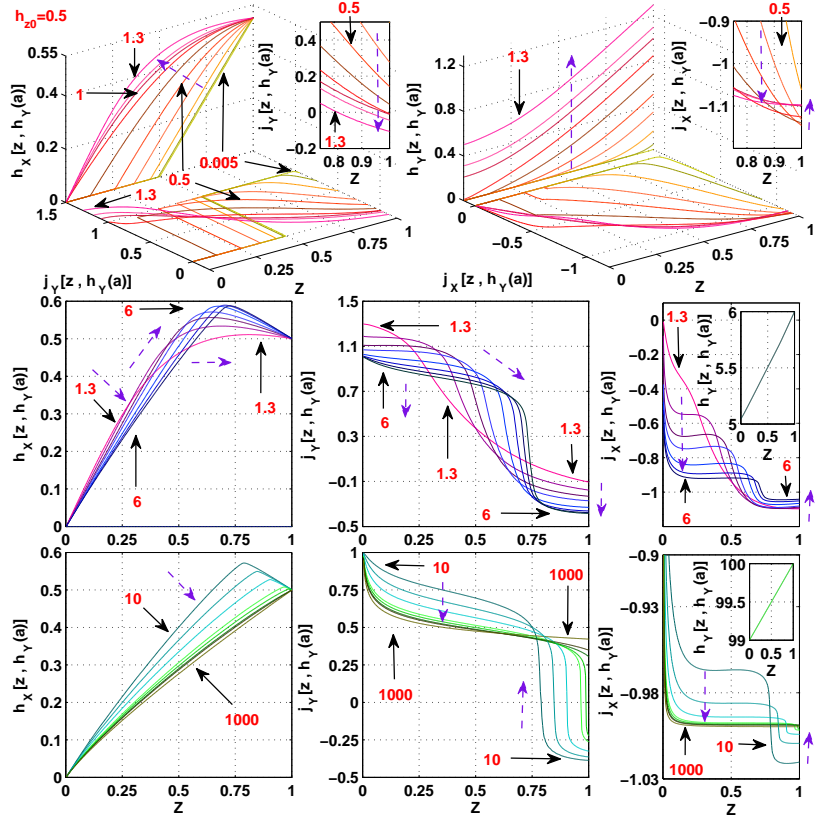}}
\caption{(Color online) Profiles of the magnetic field components
$h_{x}[z,h_{y}(a)]$ and $h_{y}[z,h_{y}(a)]$ for a perpendicular field component
$h_{z0}=0.5$. Also included are the corresponding current-density profiles
$j_{y}[z,h_{y}(a)]$ and  $j_{x}[z,h_{y}(a)]$ for the CT1 model. For clarity, the
longitudinal magnetic field is applied in three stages: (Top) $h_{y}(a)=0.005,
0.050, 0.170, 0.340, 0.500, 0.680, 0.845, 1.0, 1.1$, $1.3$, (Middle)
$h_{y}(a)=1.3, 1.6, 1.9, 2.2, 3.0, 4.0, 5.0, 6.0$, and (Bottom) $h_{y}(a)=10,
20, 40, 80, 100, 125, 150, 1000$.}
\label{Fig_5}
\end{center}
\end{figure}
In this section, we show the results obtained for the {\em square} condition
given by $\chi\equiv j_{c\parallel} / j_{c\perp} = 1$ (CT1 in what follows).
This can be considered as a lower bound for such quantity because the
experimental values reported in the literature are typically above unity.

\begin{figure}[h]
\begin{center}
{\includegraphics[width=0.55\textwidth]{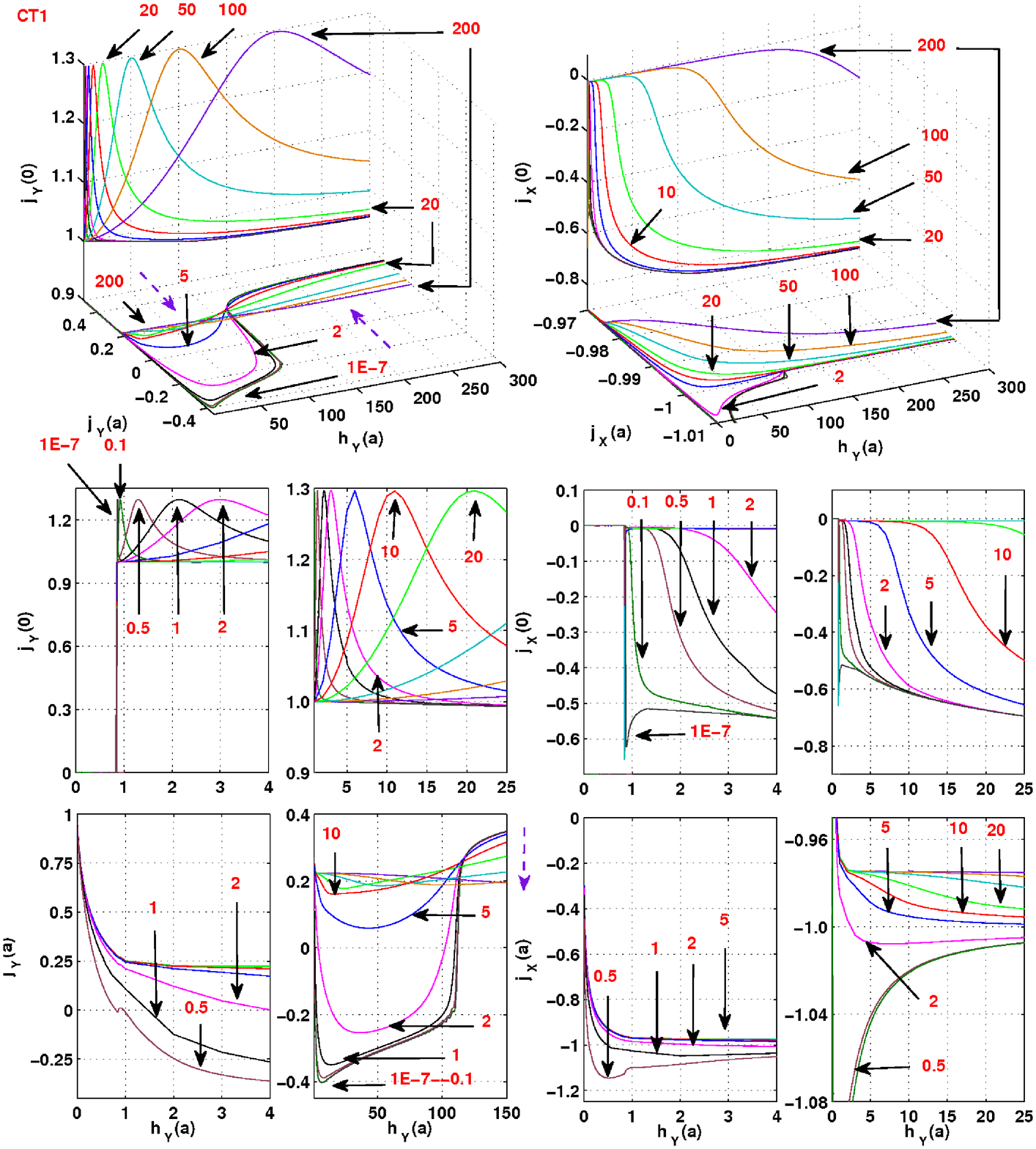}}
\caption{(Color online) Dynamics of the current density vector for the CT1
model. To the left, we show the patterns of transport current over the central
layer ($j_{y}(z=0)$) and external layer ($j_{y}(z=a)$) of the slab. To the
right, we show the variation of the perpendicular component $j_{x}(z=0)$ and
$j_{x}(z=a)$.  The curves are labelled according to the perpendicular magnetic
field component $h_{z0}=0.1, 0.5, 1, 2, 5, 10, 20, 50, 100, 200$.}
\label{Fig_7}
\end{center}
\end{figure}

In order to obtain continuity with the T-state results obtained above, the
electrodynamic quantities of interest have been proceeded under the same
arguments developed in Sec.~\ref{SubSection_2_2} as regards to the magnetic
process shown at Fig.~\ref{Fig_1}c. On the other hand, with the aim of getting a
detailed physical interpretation on how the longitudinal and transverse magnetic
fields affect the dynamics of the transport current problem, we show the
magnetic penetration profiles for low and high perpendicular fields, i.e.,
$h_{z0}=0.5$ (Fig.~\ref{Fig_5}) and $h_{z0}=10$ (Fig.~\ref{Fig_6}). Eventually,
for completeness, the set of initial conditions $h_{z0}$ is extended in
(Fig.~\ref{Fig_7}). It will be shown that the fingerprint of the CT model is
identified as a peak effect in the magnetization curves (Fig.~\ref{Fig_8})
caused by the maximal enhancement of the critical current transport density
along the central layer.

In order to ease the interpretation of the intricate behavior of the magnetic
profiles under CT conditions, we have divided the experimental process in three
successive stages as the longitudinal magnetic field component $h_{y}(a)$ is
increased, i.e.: 
%
%
\begin{figure}
\begin{center}
{\includegraphics[width=0.48\textwidth]{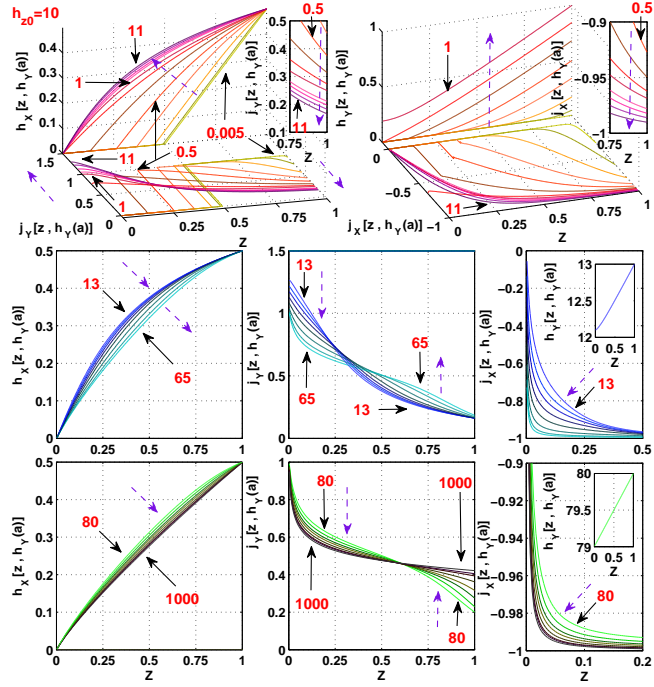}}
\caption{(Color online) Same as Fig.\ref{Fig_5}, but for $h_{z0}=10$ and the
values of the longitudinal field: (Top) $h_{y}(a)=0.005, 0.050, 0.170, 0.340,
0.500, 0.680, 0.845, 1.0, 4.0$, $6.0, 8.0, 11$, (Middle) $h_{y}(a)=13, 15, 17,
20, 25, 35, 50, 65$, (Bottom) $h_{y}(a)=80, 100, 125, 150, 200, 300, 400,
1000$.}
\label{Fig_6}
\end{center}
\end{figure}
%
%

\begin{itemize}
\item[(\textit{i})] The current density at the center $j_{y}(0)$ increases until
a maximum value is obtained (top of Figs.~\ref{Fig_5} and \ref{Fig_6}). Notice
that the partial penetration regime is included within this stage .
\item[(\textit{ii})] The minimum value for the longitudinal current density
along the superconducting surface is reached (middle row of Figs.~\ref{Fig_5}
and \ref{Fig_6}).
\item[(\textit{iii})] The longitudinal current density $j_{y}(a)$ stabilizes
around $j_{y}(a)\approx0.5$ (Bottom of Figs.~\ref{Fig_5} and \ref{Fig_6}).
\end{itemize}

\subsubsection{Field and current density penetration profiles}
To start with, notice that the trend of the profiles for the partial penetration
regime is quite independent of the perpendicular magnetic field $h_{z0}$
(Figs.~\ref{Fig_5} and \ref{Fig_6}). Moreover, the partial penetration regime in
which the transport current zone progressively penetrates the sample is
practically independent of the magnetic anisotropy of the critical state
(compare to Fig.~\ref{Fig_2}). On the other hand, the negative current patterns
are also found under the low applied magnetic fields $h_{z0}<0.5$. However, by
contrast to the results within the previous section, recall that for the T-state
model the condition $j_{c ||}\rightarrow\infty$ allows unbounded values for the
longitudinal current $j_{y}$ at the center of the sample. By contrast, for the
bounded case, the magnetic anisotropy of the material law $\Delta_{r}$ defines
the maximal current density for the critical state regime. In other words, the
maximal length of the vector $\textbf{J}$ within the region $\Delta_{r}$ defines
the maximal transport current allowed in the superconductor.

%
%
%
%

Recalling Ref.\cite{Badia_2009,Ruiz_2010},  the maximal longitudinal transport
current density corresponds to the optimal orientation in which the biggest
distance within the superelliptic region is obtained, i.e.:
%
%
\begin{eqnarray}
j_{y}^{max}=\left( 1+\chi^{2n/(n-1)} \right)^{(n-1)/2n}
 \, .
\label{Eq_19}
\end{eqnarray}
Note that his equation allows to obtain the maximum value expected for $j_{y}$
in terms of the actual critical state model in use. Thus, for $\chi = 1, n=4$
one gets $j_{y}^{max}=1.3$, a value that can be checked in Figs.~\ref{Fig_5} and
\ref{Fig_6}.
\subsubsection{Analysis of the current density behavior}
Complimentary to the field penetration profiles, Fig.~\ref{Fig_7} shows the
dynamics of the main components of the current density along the central and
external layers of the superconducting slab in the CT1 condition. Once again,
notice that the full penetration regime can be clearly distinguished from the
partial penetration regime.

Also, an interesting property to be noticed is that the value $j_{y}^{max}(0)$
is independent of the applied magnetic field at least as regards the existence
of the peak effect in the transport current density (Fig.~\ref{Fig_7}). Thus,
the enhancement of the transport current density can be either accelerated or
decelerated with the tilt of the applied magnetic field, but in general terms,
its maximum directly relates to the limitation introduced by the cutting
mechanism. Physically, this means that the role played by the magnetic
anisotropy of the material law may be characterized by the influence of the
threshold cutting value on the enhancement of the critical current density. This
point will be made clearer when material laws with different values of $\chi$
are considered. 
%
%
\begin{figure}[h]
\begin{center}
{\includegraphics[width=0.5\textwidth]{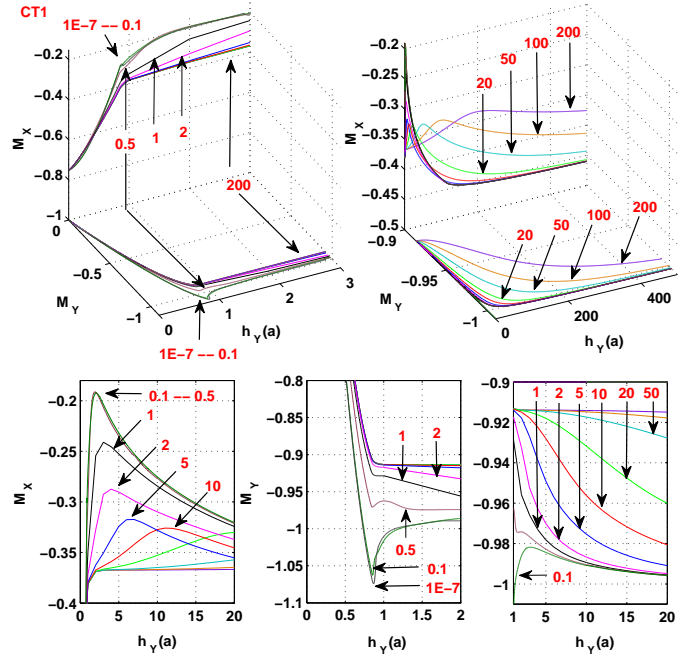}}
\caption{(Color online) Magnetic moment $(M_{x},M_{y})$ of the slab as a
function of the applied magnetic field component $h_{y}(a)$ for the CT1 model.
The curves are labelled according to the perpendicular magnetic field component
$h_{z0}$ for each.}
\label{Fig_8}
\end{center}
\end{figure}
%
%
\subsubsection{Magnetic moment of the sample}
Finally, notice that the limitation introduced by the flux cutting mechanism
imposes a maximal compression of the current density within the sample. Thus,
the peak effects both for the transport current density $j_{y}$
(Fig.~\ref{Fig_7}) and for the magnetic moment component $M_{x}$
(Fig.~\ref{Fig_8}) are defined by the instant at which the maximal transport
current density occurs. Additionally, upon further increasing the longitudinal
applied magnetic field component $h_{y}(a)$, the profile $h_{x}(z)$ will be
forced to decrease from the central sheet ($z=0$) towards the external surface
($z=a$). This reversal generates a local distortion of the longitudinal current
density $j_{y}$ in a bow tie shape (see the middle row of Figs.~\ref{Fig_5} and
~\ref{Fig_6}). Likewise, as soon as the profile $j_{||}(0)=j_{c||}$) is reached,
the magnetic moment $M_{x}$ starts decreasing (Fig.~\ref{Fig_8}) as one can see
by comparison of Figs. \ref{Fig_7} and \ref{Fig_8}.
%
%
\subsection{CT$\chi$-states: Emergence of the cutting
threshold}~\label{SubSection_3_3}
%
%
\begin{figure*}[h]
\begin{center}
{\includegraphics[width=0.98\textwidth]{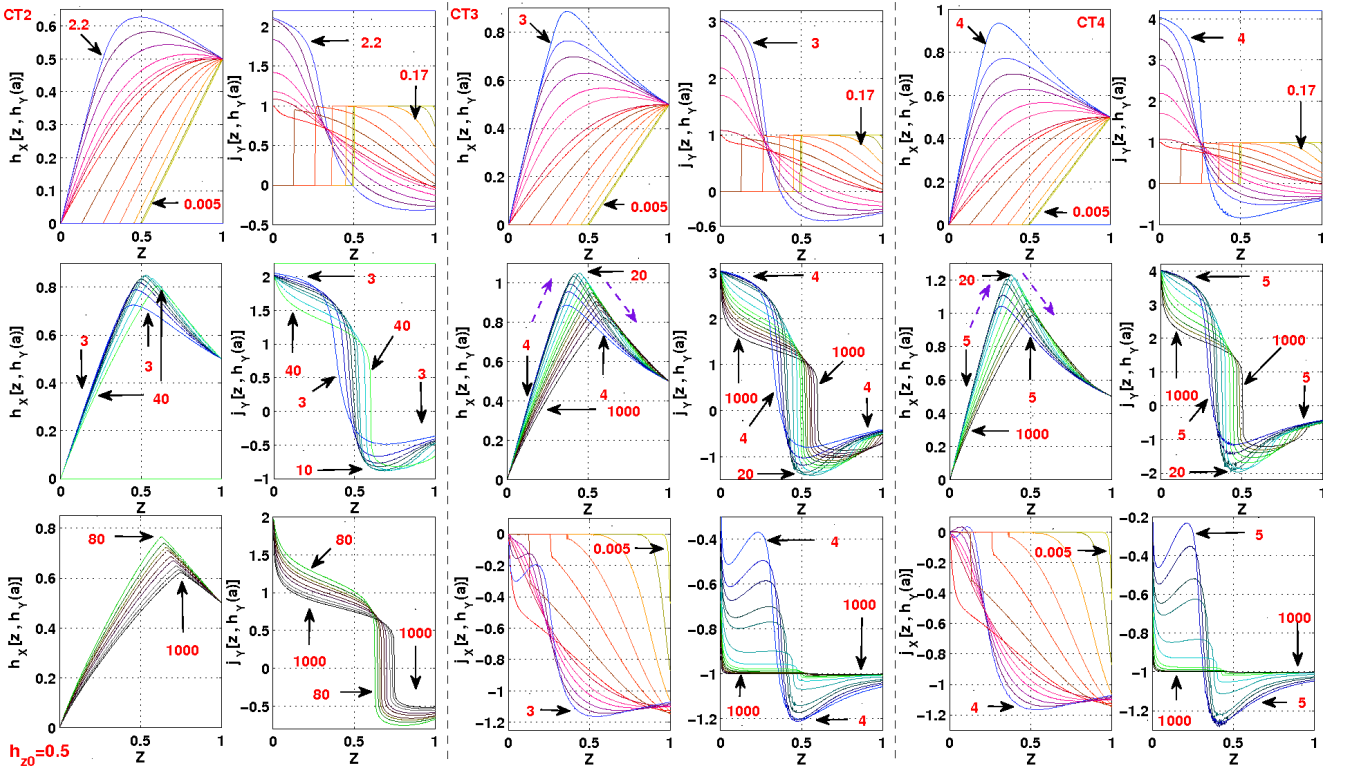}}
\caption{(Color online) Profiles of $h_{x}[z,h_{y}(a)]$ and $j_{y}[z,h_{y}(a)]$
for the 1st (top) and 2nd (middle) stages of the magnetic dynamics described in
CT2, CT3, and CT4 cases (see text) all under a field $h_{z0}=0.5$. The 3rd stage
is only defined for the CT2 case (bottom). $h_{y}(a)$ is as follows. CT2: 1st
stage $h_{y}(a)=0.005, 0.050, 0.170, 0.340, 0.500, 0.680, 0.845, 1.0, 1.1$,
$1.3, 1.6, 1.9, 2.2$, 2nd stage $h_{y}(a)= 3, 4, 5, 6, 8, 10, 20, 40$, and 3rd
stage $h_{y}(a)= 80, 120, 160, 200, 300, 400, 600, 800, 1000$. CT3: 1st stage
$h_{y}(a)=0.005, 0.050, 0.170, 0.340, 0.500, 0.680, 0.845, 1.0, 1.5$, $1.8, 2.2,
2.6, 3.0$, and 2nd stage $h_{y}(a)=4, 5, 6, 8, 15, 20, 40, 70, 100, 150, 200,
300, 400, 600, 1000$. CT4: 1st stage $h_{y}(a)=0.005, 0.050, 0.170, 0.340,
0.500, 0.680, 0.845, 1.0, 1.5$, $1.8, 2.2, 2.6, 3.0, 4.0$, and 2nd stage
$h_{y}(a)=5, 6, 8, 15, 20, 40, 100, 200, 400, 600, 1000$. } 
\label{Fig_9}
\end{center}
\end{figure*}
%

The intrinsic interplay between the cutting and depinning mechanisms introduces
difficulties when one tries to extract the threshold values for a typical
experimental arrangement. Nevertheless, as it was argued in the previous
section, several fingerprints of the actual material law can be identified,
i.e.: the longitudinal current density along the central and surface
superconducting layers, and the behavior of the magnetic moment curves. In order
to complement this scenario, we have simulated three additional experiments,
defined by the so-called CT2, CT3, and CT4 conditions, in which the threshold
value of the cutting current density is varied relative to the depinning limit
($\chi = 2,3,4$ respectively). 
%
%
\begin{figure*}
\begin{center}
{\includegraphics[width=0.98\textwidth]{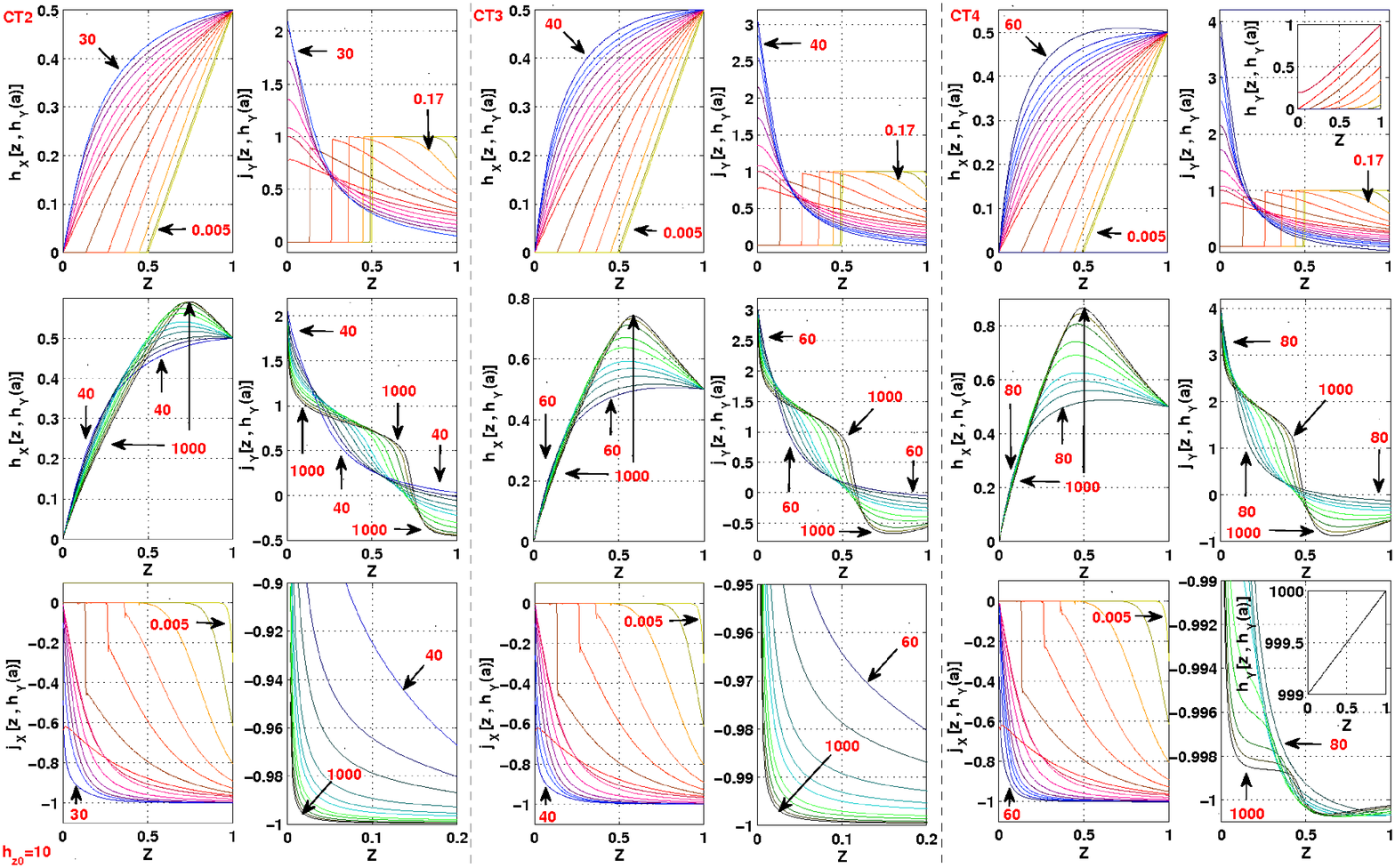}}
\caption{(Color online) Similar to Fig.~\ref{Fig_9}, but for a perpendicular
magnetic field component $h_{z0}=10$. The curves are labelled as follows. (CT2 -
left panel): 1st stage  $h_{y}(a)=0.005, 0.050, 0.170, 0.340, 0.500, 0.680,
0.845, 1.0, 5.0$, $10, 15, 20, 30$, and 2nd stage $h_{y}(a)=40, 60, 80, 120,
160, 200, 300, 400, 600, 800, 1000$. (CT3-middle panel): 1st stage
$h_{y}(a)=0.005, 0.050, 0.170, 0.340, 0.500, 0.680, 0.845, 1.0, 5.0$, $10, 15,
20, 25, 30, 40$, and 2nd stage $h_{y}(a)=60, 80, 120, 160, 200, 300, 400, 600,
800, 1000$. (CT4 - right panel): 1st stage $h_{y}(a)=0.005, 0.050, 0.170, 0.340,
0.500, 0.680, 0.845, 1.0, 5.0$, $10, 15, 20, 35, 30, 40, 60$, and 2nd stage
$h_{y}(a)=80, 120, 160, 200, 300, 400, 600, 800, 1000$.}
\label{Fig_10}
\end{center}
\end{figure*}

\subsubsection{Field and current density penetration profiles}
The behavior of the local electrodynamic quantities is displayed in terms of the
aforementioned three stages. In order to allow comparison, the main features
displayed in Figs.~\ref{Fig_5} and \ref{Fig_6} for the CT1 condition with
$h_{z0}=0.5$ and 10, are also shown in Figs.~\ref{Fig_9} and \ref{Fig_10} under
the conditions CT2, CT3 and CT4. Several peculiarities are to be noted. To start
with, the emergence of negative states for the transport current density close
to the external surface of the superconducting sample will be more evident
either when $h_{z0}$ is reduced and/or the current density anisotropy factor
$\chi$ is increased.

%

%

%
\begin{figure*}
\begin{center}
{\includegraphics[width=0.98\textwidth]{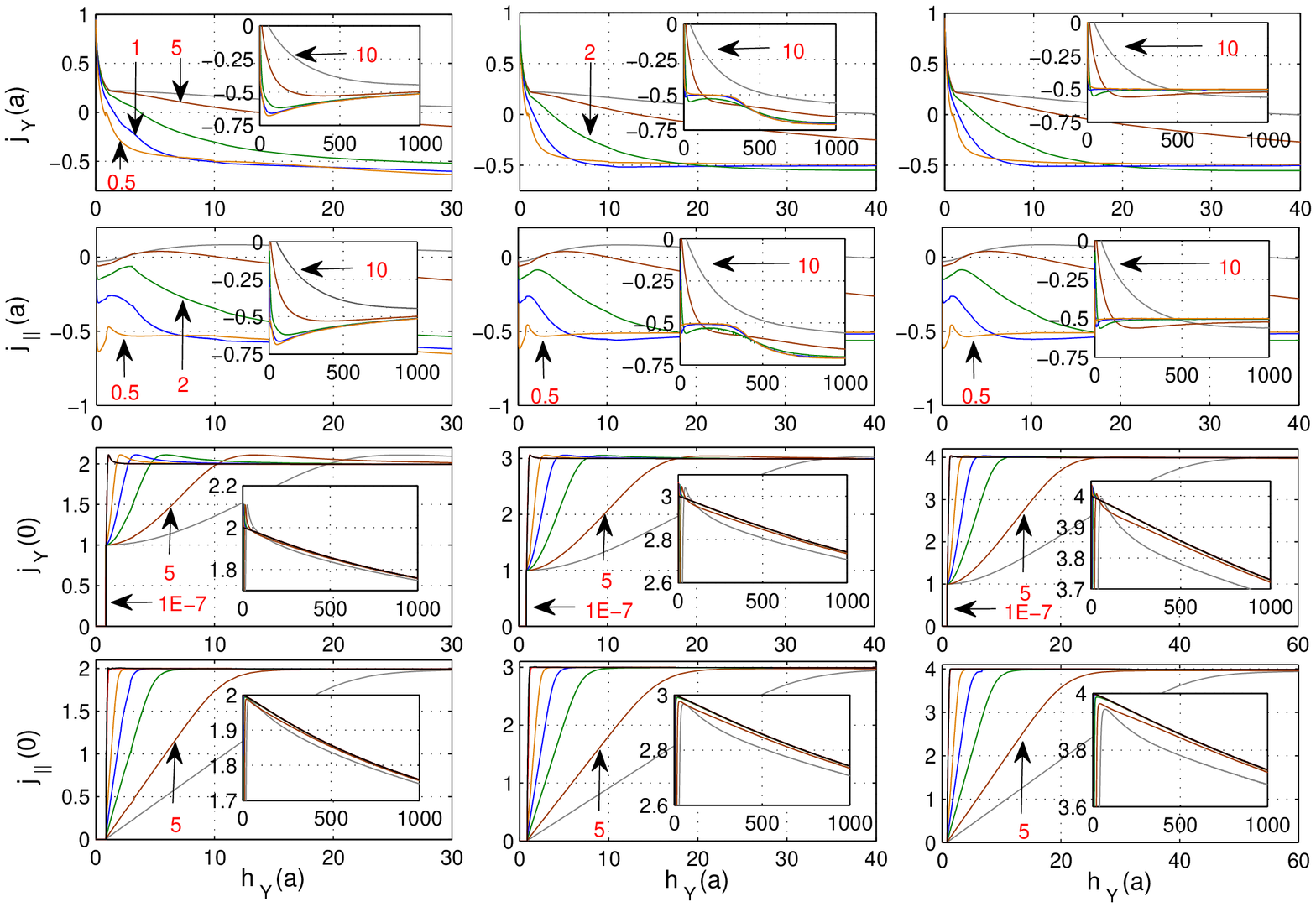}}
\caption{(Color online) Dynamics of the current density vector for the CT2, CT3
and CT4 models. In the upper panel, we show the patterns of transport current
along the central ($j_{y}(z=0)$) and surface layers ($j_{y}(z=a)$) of the slab.
The curves are labelled according to the perpendicular magnetic field component
$h_{z0}=1E-7, 0.5, 1, 2, 5, 10$}
\label{Fig_11}
\end{center}
\end{figure*}
%
%

\subsubsection{Analysis of the current density behavior}
In order to confirm the above interpretation, we show the magnetic dynamics of
the longitudinal current density $j_{y}$, the cutting current component $j_{||}$
 (Fig.~\ref{Fig_11}) for the conditions CT2, CT3, and CT4. We have taken a wide
set of values for the perpendicular field component ($h_{z0}$). 
On the one hand, as regards the sample's surface, Fig.~\ref{Fig_11} shows that
the longitudinal current density $j_{y}(a)$ does not display significative
differences when one has $\chi\geq2$ (see also Fig.~\ref{Fig_3} for
$\chi=\infty$). 
Thus, the disappearance of the negative current flow along the external
superconducting surface does not occur despite a very high applied magnetic
field has been considered ($h_{y}(a)=1000$). On the other hand, it is important
to notice that the patterns of the parallel current density along the
superconducting surface ($j_{||}(a)$) are almost indistinguishable as soon as
the condition $\chi\geq 2$ (CT2) is reached (upper half of Fig.~\ref{Fig_11}).
This implies that for an accurate picture of the parallel critical current,
surface properties do not provide a useful information.

However, outstandingly, Fig.~\ref{Fig_11} shows that the threshold value for the
cutting current density can be estimated from the experimental measurement of
the transport current density along the central sheet of the superconducting
sample. 
Notice further that, regardless of the experimental conditions ($h_{z0}$,
$h_{y}(a)$) and also for different bandwidths $\chi$ no significant change
occurs in the parallel current density around the central sheet of the sample
(lower half of Fig.~\ref{Fig_11}). 

\subsubsection{Magnetic moment of the sample}
\begin{figure*}
\begin{center}
{\includegraphics[width=0.98\textwidth]{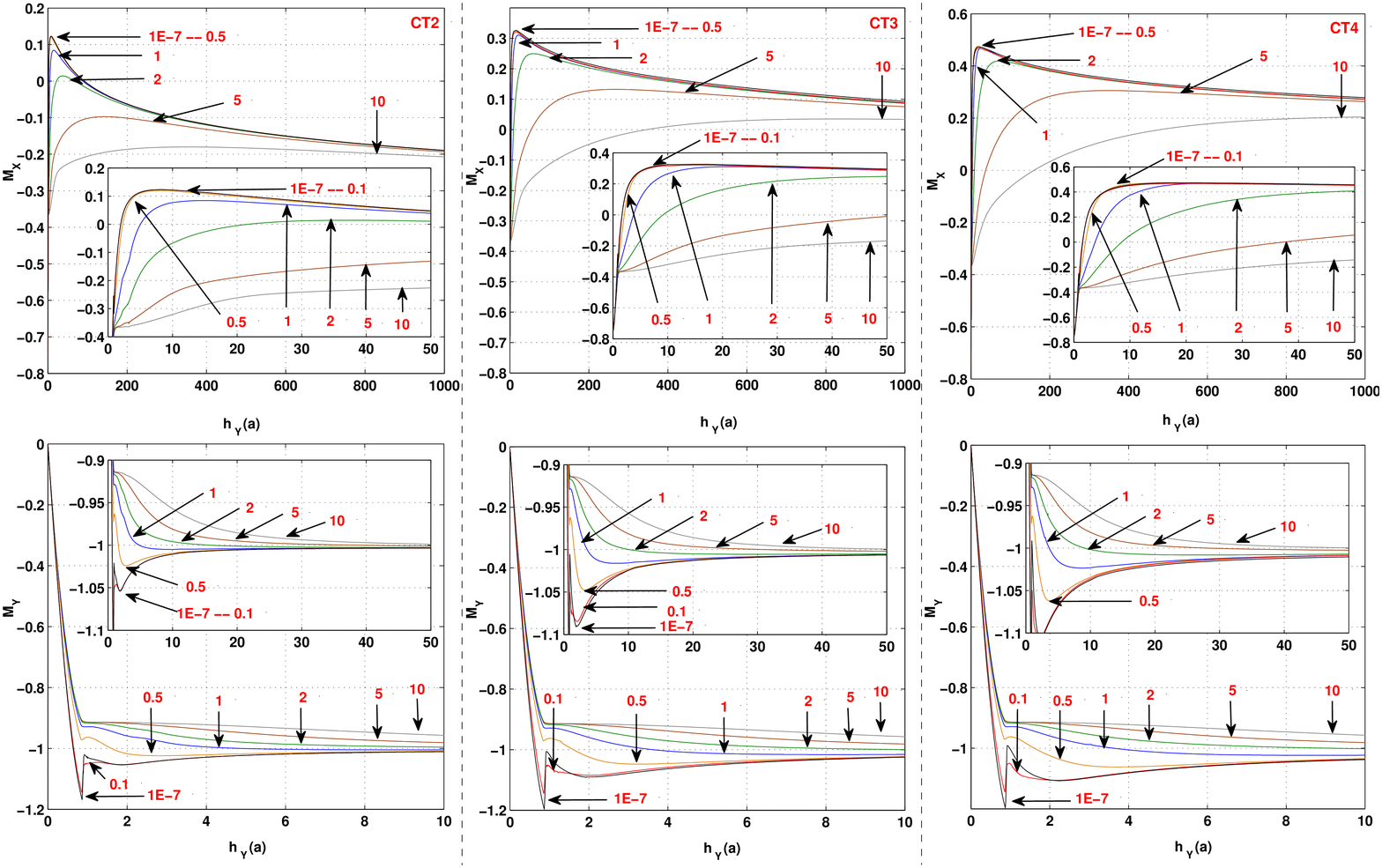}}
\caption{(Color online) The magnetic moment $(M_{x},M_{y})$ of the slab as a
function of the applied magnetic field component $h_{y}(a)$ for the CT2, CT3,
and CT4 smooth critical state models. The curves are labelled according to the
perpendicular magnetic field component $h_{z0}$ in correspondence at the
Fig~\ref{Fig_11}.}
\label{Fig_12}
\end{center}
\end{figure*}

The peak effects observed both in the local transport current density $j_{y}(0)$
(lower half of Fig.~\ref{Fig_11}), and on the global magnetic moment $M_{x}$
(Fig.~\ref{Fig_12}), are predicted to appear subsequent to the maximal
longitudinal transport current density $j_{y}^{max}$ (Eq.~\ref{Eq_19}). 

%
%
%
%

Furthermore, one additional feature is to be noted: the interval between the
instant at which the maximal transport current condition is reached
($j_{y}(0)=j_{y}^{max}$), and the instant at which the slope of the magnetic
moment $M_{x}$ changes sign, could be assumed as transient or stabilization
period required for an accurate determination of the value $j_{c ||}$ when
measurements are performed in terms of the applied longitudinal field $h_{y}$.
Apparently, this transient increases with the value of the perpendicular field
$h_{z0}$. From this point on, $j_{y}(0)$ may be basically identified with
$j_{c\|}$.


\section{Conclusions}~\label{Section_4}
Despite of extensive experimental and theoretical studies about the
electrodynamic response of type-II superconductors in longitudinal geometries,
much uncertainties remain about the interaction between flux depinning and
cutting mechanisms, and their influence in such striking observations as the
appearance of negative transport current flow, the enhancement of the critical
transport current density, and the observation of peak effects on the
magnetization curves. In this article, and based on the application of our
general critical state theory, we have reproduced theoretically the existence of
negative flow domains, local and global paramagnetic structures, emergence of
peak-like structures in the longitudinal magnetic moment, as well as the
compression of the transport current density for a wide number of experimental
conditions.

The longitudinal transport problem in superconducting slab geometry has been
studied as follows: we have considered a superconducting slab lying at the $xy$
plane and subjected to a transport current density along the $y$ direction. The
slab is assumed to be penetrated by a uniform vortex array along the $z$
direction, so that the local current density along the thickness of the sample
is entirely governed by the depinning component $J_{\perp}$, perpendicular to
the local magnetic field. Subsequently, a magnetic field source parallel to the
transport current direction is switched on. Then, the experimental conditions
have been modulated through the value of the external magnetic field $H_{y0}$.
The dynamical behavior of the transport current density is shown to rely on the
interaction between the cutting and depinning mechanisms. Moreover, the
intensity of the inherent effects has been shown to depend on the perpendicular
component $H_{z0}$, being more prominent as this quantity is reduced. By means
of our general critical state theory that allows to modulate the influence of
the different physical effects, we  have been able to show that the peak
structures observed in the magnetization curves and the patterns of the
transport current along the central section of a superconducting sample are both
directly associated with the local structure of the vortex lattice. Such
dependence may become more pronounced as the extrinsic pinning of the material
is reduced, in favor of the flux cutting interactions. 
The same conclusion was pointed out from the experimental measurements of
Blamire et. al (Ref.~\cite{Blamire_2003,Blamire_1986}) for high critical
temperature and conventional superconductors. By using a theory that depends on
two parameters ($\chi\equiv J_{c\|}/J_{c\perp}$ and $n$) accounting respectively
for the intrinsic material anisotropy and for the smoothness of the
$J_{\|}(J_{\perp})$ law, here, we have quantitatively investigated the influence
of the flux cutting mechanism. It has been done by comparing the so-called
square model ($\chi=1$), the T-state model ($\chi\to\infty$), and the double
critical state conditions CT$\chi$ with $\chi=2$, $\chi=3$, and $\chi=4$, all of
them with the {\em smoothing} index $n=4$. 

From our theoretical framework we obtain that the isotropic model (circular
region: $\chi=1, n=1$) does neither predict the appearance of negative current
patterns nor the peak effects in the magnetic moment curves. However, as long as
a clear distinction between the depinning and the cutting components of $\bf J$
is allowed, several remarkable facts can be explained.

Going into detail, when the cutting threshold is high ($J_{c\|}\gg J_{c\perp}$
or $\chi \gg 1$) the emergence of negative current patterns is ensured because
unbounded parallel current density allows unconstrained rotations for the flux
lines as the longitudinal magnetic field increases. Thus, under a range of
conditions, the peak effects in the magnetic moment and a modulation of the
negative surface currents are predicted. 

Concentrating on the local properties within the sample, a clear independence of
the field penetration profiles relative to the anisotropy level of the material
law has been obtained for the partial penetration regime. On the other hand, as
soon as the full penetration state is reached, noticeable effects of the
magnetic anisotropy law are predicted both within the central and external
layers of the superconducting sample. Thus, according to our results (see
Fig.\ref{Fig_11}) the {\em elusive} parallel critical current density parameter
$J_{c||}$  can be obtained from the local measurement of the transport current
density along the central layer of the superconducting sample.


\section*{Acknowledgement}

This work was supported by Spanish CICyT and FEDER program (project
MAT2008-05983-C03-01) and by the DGA grant T12/2011. H. S. Ruiz acknowledges the
financial support provided by the Spanish CSIC JAE-program.

%
%
\section*{References}

\end{document}